\documentclass[sigconf,authorversion,nonacm]{acmart}
\usepackage{multicol}
\usepackage{multirow}
\usepackage{makecell}
\usepackage{longtable}
\usepackage{adjustbox}
\usepackage{caption}
\usepackage{subcaption}
\usepackage{graphicx}
\usepackage{svg}
\usepackage{amsmath}
\usepackage{hyperref}
\usepackage{url}
\usepackage{tablefootnote}

\usepackage{epsf} 

\usepackage[flushleft]{threeparttable}
\usepackage{tcolorbox}

\usepackage{titlesec}
\begin{document}

\title{Guiding the retraining of convolutional neural networks against adversarial inputs}%: best guidance metrics and configurations}

\author{Francisco Durán}
\email{francisco.javier.duran.lopez@upc.edu}
\orcid{}
\affiliation{
  \institution{Universitat Politècnica de Catalunya - BarcelonaTech}
  \streetaddress{Jordi Girona, 1-3}
  \city{Barcelona}
  \state{Catalunya}
  \country{Spain}
  \postcode{08034}
}
\affiliation{
  \institution{Universidad Nacional Autónoma de México}
  \streetaddress{ 	Av Universidad 3000, Ciudad Universitaria}
  \city{Mexico City}
  \state{}
  \country{Mexico}
  \postcode{04510}
}

\author{Silverio Martínez-Fernández}
\email{silverio.martinez@upc.edu}
\orcid{0000-0001-9928-133X}
\affiliation{
  \institution{Universitat Politècnica de Catalunya - BarcelonaTech}
  \streetaddress{Jordi Girona, 1-3, omega oficina 004}
  \city{Barcelona}
  \state{Catalunya}
  \country{Spain}
  \postcode{08034}
}

\author{Michael Felderer}
\email{michael.felderer@uibk.ac.at}
\orcid{0000-0003-3818-4442}
\affiliation{
  \institution{University of Innsbruck}
  \streetaddress{Technikerstr. 21a}
  \city{Innsbruck}
  \state{Tyrol}
  \country{Austria}
  \postcode{6020}
}

\author{Xavier Franch}
\email{xavier.franch@upc.edu}
\orcid{}
\affiliation{
  \institution{Universitat Politècnica de Catalunya - BarcelonaTech}
  \streetaddress{Jordi Girona, 1-3}
  \city{Barcelona}
  \state{Catalunya}
  \country{Spain}
  \postcode{08034}
}

\renewcommand{\shortauthors}{Durán et al.}

\begin{abstract}
Background: When using deep learning models, there are many possible vulnerabilities and some of the most worrying are the adversarial inputs, which can cause wrong decisions (e.g., wrongly classifying an image) with minor perturbations. Therefore, it becomes necessary to retrain these models against adversarial inputs, as part of the software testing process addressing the vulnerability to adversarial inputs. Furthermore, for an energy efficient testing and retraining process, data scientists need support on which are the best guidance metrics for reducing the adversarial inputs to use as well as optimal dataset configurations. 

Aims: We examined four guidance metrics for retraining deep learning models, specifically with convolutional neural network architecture, and three retraining configurations. Our goal is to improve the convolutional neural networks against adversarial inputs with regard to accuracy, resource utilization and time from the point of view of a data scientist in the context of image classification.

Method: We conduced an empirical study in two datasets for image classification. We explore: (a) the accuracy, resource utilization and time of retraining convolutional neural networks by ordering new training set by four different guidance metrics (neuron coverage, likelihood-based surprise adequacy, distance-based surprise adequacy and random), (b) the accuracy and resource utilization of retraining convolutional neural networks with three different configurations (from scratch and augmented dataset, using weights and augmented dataset, and using weights and only adversarial inputs).

Results: We reveal that retraining with adversarial inputs from original model weights and by ordering with surprise adequacy metrics gives the best model w.r.t. accuracy, resource utilization and time.

Conclusions: Although more studies are necessary, we recommend data scientists to use the above configuration and metrics to deal with the vulnerability to adversarial inputs of deep learning models, as they can improve their models against adversarial inputs without using many inputs. We also show that dataset size has an important impact on the results.

\end{abstract}



\maketitle

%\begin{multicols}{2}

\section{Introduction}
In recent years, Deep Learning (DL) systems, defined as those software systems with functionalities enabled by at least one DL model, have become a widespread machine learning approach, due to their outstanding capacity to solve complex problems. The use of DL systems ranges from applications in autonomous driving systems to applications in medical treatments, among others \cite{stallkamp2012man,castanyer2021integration,tran2021recommender}. However, given their very nature, it is often the case that DL systems exhibit an unexpected behavior or produce anomalous results when inferring with inputs differents to those used in the training stage, due to their statistical and black box nature. These vulnerabilities or system errors can lead to undesirable consequences when integrated in real-world applications already in production.

The existence of vulnerabilities is commonplace  in any type of software system. In traditional systems, i.e., not embedding any DL system, we may find a vast amount of testing approaches to ensure a certain level of robustness. When the tests detect a problem, there are specific solutions to fix the problem or at least mitigate the effects to a large extent \cite{khan2011different}. In contrast, testing DL systems is completely different, due to their non-deterministic nature: the ``correct'' answer or ``expected'' behaviour in response to a certain input is often unknown at the time of training the DL model \cite{barr2014oracle,amershi2019software}. 

Recently, one of the most worrying vulnerabilities in DL systems are adversarial inputs, which are inputs created with slight modifications to the original inputs of a dataset. These types of inputs are often not distinguishable by the human from the input from which they are generated, but they lead the DL system to produce an incorrect output.

Given this scenario, DL system testing has become a relatively new broad research field, in which every year the number of publications related to the subject increases \cite{zhang2019machine,huang2020survey,martinez2021software}, aiming at laying the foundations of the tests to be able to compare research results in a consistent and effective way. Different frameworks have been created to facilitate the use of these methods \cite{papernot2016technical,rauber2017foolbox,zhang2018deeproad,gerasimou2020importance,ahuja2022testing} and likewise a number of metrics have been created to compare the properties of these testing methods \cite{pei2017deepxplore, ma2018deepgauge, kim2019guiding}.

This work focuses on improving the testing and retraining of DL models in the presence of adversarial inputs. To this end, we apply two different approaches: \emph{(i)} using the information provided by guidance metrics, which try to identify the most useful inputs according to the criteria of each metric, and  \emph{(ii)} adding adversarial inputs to augment the dataset employing different training inputs from the original training set. Still, questions arise: When and how to use a certain metric? What is the accuracy, resource utilization and time of using such metrics? When and how to use a certain retraining configuration? What is the accuracy and resource utilization of using a certain retraining configuration? In this context, \textbf{this work aims at comparing both guidance metrics and configurations for retraining DL models with adversarial inputs based on their accuracy, resource utilization and time}. The results of the research benefit data scientists to test their models' accuracy against adversarial inputs while keeping the trade-off with the resource utilization of the retraining process. We evaluate our method in one typical DL field application, namely image classification. For this reason, we focus on one particular type of DL model, namely Convolutional Neural Networks (CNN), which are widely used for image classification due to their high performance. Since LeCun \emph{et al.} introduced them \cite{lecun1989backpropagation}, there has been significant progress in the literature with respect to their accuracy even in challenging datasets, as well as the explanation of their behaviour and abstraction \cite{krizhevsky2012imagenet,zeiler2014visualizing}.

The \textbf{main contributions} of this work are:
\begin{itemize}
\item A comparison, in terms of accuracy, resource utilization and time, of four guidance metrics, including random selection, and three configurations of retraining CNN with adversarial inputs.
\item A replication package, available online. \footnote{Please refer to \url{https://doi.org/10.5281/zenodo.5904550}}
\end{itemize}

This document is structured as follows. Section \ref{sec:background} respectively introduces the DL arquitecture used for the models, the guidance metrics (Neuron Coverage (NC) and two Surprise Adequacy (SA) metrics), and the adversarial attack Fast Gradient Sign Method (FGSM) used to create the adversarial inputs. Section \ref{sec:related work} describes related work. Section \ref{sec:empirical_study} presents the research questions and the study design. Section \ref{sec:results} presents the results. Section \ref{sec:discussions} presents the discussions. Section \ref{sec:threats} describes the threats to validity, and Section \ref{sec:conclusions} draws conclusions and future work.

\begin{table*}[t]
    \centering
        \caption{Review of the retraining configurations used in related work (C$k$ refer to the configurations presented in Section \ref{sec:empirical_study})}
        \label{tab:review_retrainings}
    \begin{tabular}{cp{4.5cm}p{3.5cm}p{3cm}p{3cm}}
    \toprule
    Study & Configuration & Independent variables & Studied variables & Datasets \\
    \midrule
        
    Pein \emph{et al.}\cite{pei2017deepxplore} &
    
    Retraining of an original model with 100 new error-inducing samples (Similar to C3).&
    
    DL model&

    Accuracy&
    MNIST
    
    ImageNet
    
    Udacity challenge
    
    Contagio/VirusTotal
    
    Drebin
    \\
    
    \hline
    
    Tian \emph{et al.}\cite{tian2018icse} &
    
    Retraining using synthetic images and original training set. Does not specify if it is done from scratch (Similar to C1 and C2).&
    
    Image transformation&
    
    Accuracy
    
    MSE&
    Udacity challenge
    \\
    
    \hline
    
    Kim \emph{et al.}\cite{kim2019guiding} &
    
    Retraining with 100 new inputs from different ranges of SA metrics (Similar to C3).&
    
    Layer of neurons used for SA computation
    
    Adversarial attack&
    
    Accuracy
    
    MSE&
    MNIST
    
    CIFAR-10
    
    Udacity challenge
    \\
    
    \hline
    
    Ma \emph{et al.}\cite{ma2021test} &
    
    As C3, however on each iteration they compute the metrics, instead of calculating metrics once.&

    Testing metric&

    Accuracy
    
    Validation Loss&
    MNIST
    
    Fashion-MNIST
    
    CIFAR-10
    \\
    
    \hline
    
    Ma \emph{et al.}\cite{ma2018mode} &
    
    As C3, but when they add a  new selected batch, it starts from previous retrained model; not from original model weights.&
    
    DL model&
    
    Accuracy&
    MNIST
    
    Fashion-MNIST
    
    CIFAR-10
    \\
    
    \hline
    
    This study &
    
    C1, C2 and C3 (see Section \ref{sec:empirical_study}).&
    
    Guidance metric
    
    Retraining configuration&
    
    Accuracy
    
    Resource utilization
    
    Time&
    GTSRB
    
    Intel
    \\

        \bottomrule
        \end{tabular}
    \end{table*}
\section{Background}\label{sec:background}

In this section, we introduce: \emph{(i)} the type of DL model we used, namely Convolutional Neural Networks (CNN); \emph{(ii)} the guidance metrics used in our study, namely: Neuron Coverage (NC) and two Surprise Adequacy (SA) metrics: Likelihood-based Surprise Adequacy (LSA) and Distance-based Surprise Adequacy (DSA); \emph{(iii)} the adversarial attack used to create adversarial inputs, namely Fast Gradient Sign Method (FGSM).

\subsection{Convolutional Neural Networks (CNN)}

This is one of the most used architectures in image classification. A CNN consists of an input layer, output layer and multiple hidden layers. These hidden layers typically consist of convolutional layers, pooling layers, normalization layers, fully connected layers, or other type of layers to build more complex models. Each of the layers contains a different level of abstraction for an image dataset and the weights of the final model are obtained by backpropagation \cite{lecun1989backpropagation,lecun1998gradient,guo2017simple}.

\subsection{Guidance metrics}

Current approaches to DL systems testing evaluate them according to a number of properties, either functional (such as accuracy or precision) or non-functional (such as interpretability, robustness or efficiency) \cite{zhang2019machine}. In order to improve the DL system behaviour with respect to these properties, it is important to have metrics to compare the behaviors of the models; among the most used and accepted metrics by the community are those related to neuron coverage and to the surprise of new inputs regarding the model \cite{kim2019guiding,ma2021test,pei2017deepxplore,weiss2021review}. 

\subsubsection{Neuron Coverage (NC)}
Pei \emph{et al.} proposed the concept of NC in 2017 \cite{pei2017deepxplore} to measure the coverage of test data of a DL model and to improve the generation of new inputs, arguing that the more neurons are covered, the more network states can be explored, having a greater possibility of defect detection \cite{ma2018deepgauge}. The metric is defined as follows.
Let $D$ be a trained DL model, composed of a set $N$ of neurons. The neuron coverage of input $x$ with respect to $D$ is given by
%equation
\begin{equation}
NC(x)=\ \frac{|{\ n\ \epsilon\ N\ |\ activate(n,x)}|}{|N|}
\end{equation}
where $activate(n,x)$ is true if and only if $n$ is activated when passing $x$ to $D$.

\subsubsection{Surprise Adequacy (SA)}
Kim \emph{et al.} \cite{kim2019guiding} proposed SA as the metric that measures the degree of surprise of the model when it confronts new inputs with respect to training inputs, with the hypothesis that a good test set should be ``enough'' but not too surprising compared to the training set. Kim \emph{et al.} presented two surprise metrics, which we use in our study. 

\paragraph{Likelihood-based Surprise Adequacy (LSA).}
Let $D$ be a DL model trained on a set $T$ of inputs. The LSA of the input $x$ with respect to $D$ is given by:
%equation
\begin{equation}
LSA(x)=\frac{1}{|A_{N_L}(T_{D(x)})|}\sum_{x_t\ \epsilon\ T_{D(x)}}{K_H(\alpha_{N_L}(x)-\alpha_{N_L}(x_i))}
\end{equation}
where $\alpha_{N_L}(x)$ is the vector that stores the activation values of neurons in the $L$ layer of $D$ when $x$ is entered, $T_{D(x)}$ is the subset of $T$ composed of all the inputs of the same class as $x$, $A_{N_L}(T_{D(x)})\ =\ {\alpha_{N_L}(x_i)\ |\ x_i\ \epsilon\ T_{D(x)}}$and $K_H$ is the Gaussian kernel function with bandwidth matrix $H$ \cite{ma2021test}.

\paragraph{The Distance-based Surprise Adequacy (DSA).}
Based on the distance between vectors representing the neuronal Activation Traces of the given input and the training data (using Euclidean distance).
Let $D$ be a DL model trained on a set $T$ of inputs. The DSA of the input $x$ with respect to $D$ is given by:

%equations
\begin{equation}
    DSA\left(x\right)=\frac{\left|\left|\alpha_N\left(x\right)-\alpha_N\left(x_a\right)\right|\right|}{\left|\left|\alpha_N\left(x_a\right)-\alpha_N\left(x_b\right)\right|\right|}
\end{equation}
where:
\begin{equation}
x_a\ =\ {argmin}_{{x_i\ \epsilon\ T_{D(x)}}}\left|\left|\alpha_N\left(x\right)-\alpha_N\left(x_i\right)\right|\right|
\end{equation}
\begin{equation}
    x_b\ =\ {argmin}_{{x_j\ \epsilon\ {T \setminus T}_{D(x)}}}\left|\left|\alpha_N\left(x_a\right)-\alpha_N\left(x_i\right)\right|\right|
\end{equation}
and where $D(x)$ is the predicted class of $x$ by $D$ and $ \alpha_N\left(x\right)$ is the vector of activation values of all neurons of $D$ when confronted with $x$ \cite{ma2021test}.

\subsection{Adversarial attack: Fast Gradient Sign Method (FGSM)}
There are DL models whose dataset is increased when there is not enough data. This can be done through techniques such as obtaining new entries from those that already exist, carrying out transformations to these including cuts and rotations, among others \cite{tian2018icse,feinman2017detecting,jockel2019safe}. Another way is using adversary examples, which C. Szegedy discovered in 2013, when he noticed that several machine learning and DL models are vulnerable to slightly different inputs from those that are correctly classified, i.e. the adversarial inputs \cite{szegedy2013intriguing}. This observation caused concerns because it goes against the ability of these models to achieve a good generalization. In 2015, Goodfellow introduced the FGSM method to be able to create adversarial inputs in a relatively simple way, forcing the misclassification of the input controlling the disturbance so that it is not perceptible to a human, computed as follows \cite{goodfellow2014explaining}:
%equation
\begin{equation}
    x\ast\gets x+\varepsilon\bullet\nabla_{x}\ J(f,\theta,x)
\end{equation}
where $J$ is the cost function used for the training of the model $f$ in the neighborhood of the training point $x$ for which the adversary wants to force a wrong classification. The adversary example corresponding to the input $x$ that results from the method is denoted as $x\ast$.

\begin{table*}[]
    \captionsetup{justification=centering}
    \caption{The variables of the study}
    \label{tab:variablestab}
    \scalebox{0.94}{%
    \begin{tabular}{|l|l|l|l|l|}
        \hline
        \multicolumn{1}{|c|}{\textbf{Class}} & \multicolumn{1}{c|}{\textbf{Name}} & \multicolumn{1}{c|}{\textbf{Description}}                                                                                   & \multicolumn{1}{c|}{\textbf{Values or Formula}} & \multicolumn{1}{c|}{\textbf{Scale}} \\ \hline
        \multirow{2}{*}{Independent}         & Guidance metric                    & \begin{tabular}[c]{@{}l@{}}Testing metrics used to order the inputs for retraining\end{tabular}                          & NC, DSA, LSA, Random                                      & Nominal                             \\ \cline{2-5} 
                                             & Retraining configuration           & \begin{tabular}[c]{@{}l@{}}The configuration used for retraining the models \\ against adversarial inputs\end{tabular}    & C1, C2, C3                                    & Nominal                             \\ \hline
        \multirow{3}{*}{Dependent}           & Accuracy                           & \begin{tabular}[c]{@{}l@{}}Accuracy using the new test  set composed of the \\ original test set and an adversarial test set \\obtained from the previous applying the FGSM\\ to each one of the original test inputs\\ \\ $TP$ = True Positives; $TN$ = True Negatives\\ $FP$ = False Positives; $FN$ = False Negatives\end{tabular}  & \multicolumn{1}{c|}{$\frac{TP+TN}{TP+TN+FP+FN}$}  & Ratio                           \\ \cline{2-5} 
                                             & Resource utilization               & \begin{tabular}[c]{@{}l@{}}$u$ = Input size used to obtain the maximum accuracy\\ \\ $T$ = Number of total inputs\end{tabular} & \multicolumn{1}{c|}{$\frac{u}{T}$}                        & Ratio                           \\ \cline{2-5} 
                                             & Time                               & Time to obtain the metrics values                                                                                           & hh:mm:ss                                     & Interval                                \\ \hline
        \multirow{1}{*}{Other}               & Dataset                            & \begin{tabular}[c]{@{}l@{}}The datasets used for training the DL models\end{tabular}                                     & GTSRB, Intel                                     & Nominal                             \\ \hline
                                            
    \end{tabular}
    }
\end{table*}

\section{Related work}\label{sec:related work}
From the reviewed literature, there are many studies focused on the creation of adversarial inputs in order to uncover debilities and vulnerabilities of DL systems or models. Goodfellow \emph{et al.} \cite{goodfellow2014explaining} were the first to formulate a concrete definition of the adversarial inputs and clarify their generalization in different architectures and training sets. Additionally, they defined FGSM which allowed  to generate adversarial inputs.  Other studies centered on being able to detect such adversarial inputs for different purposes. Feinman \emph{et al.} \cite{feinman2017detecting} aimed at distinguishing adversarial samples from their normal and noisy counterparts. Ma \emph{et al.} \cite{ma2018deepgauge} used their coverage criteria to quantify defect detection ability using adversarial inputs created with different adversarial techniques. Kim \emph{et al.} \cite{kim2019guiding} used their proposed SA criteria to show that they could detect adversarial inputs.

Nevertheless, there is scarce research focused on using adversarial inputs in retraining with the purpose of improving the models \cite{kim2019guiding,pei2017deepxplore}, see Table \ref{tab:review_retrainings} for a summary. In
these few works, we can only identify one configuration used, similar to the configuration 3 presented in Section \ref{sec:empirical_study} of this work, in which the authors retrain with only a few of adversarial inputs and do not clearly show the steps for the retraining. Kim \emph{et al.} \cite{kim2019guiding} showed that sampling inputs using SA for retraining can result in higher accuracy, Pei \emph{et al.} \cite{pei2017deepxplore} showed that error-inducing inputs generated by DeepXplore can be used for retraining to improve accuracy. 

Most recently, Ma \emph{et al.} \cite{ma2021test} proposed to use testing metrics as a retraining guide, looking for an answer on how to select additional training inputs to improve the accuracy of the model. They used the original training set starting from a previous trained model but ordering these inputs following the metrics' guidance. Compared to this work, we aim to do the retraining using the information of an augmented dataset with adversarial inputs, not only the original training set.

\textbf{As the use of adversarial inputs has been proved to provide many improvements, we identified a research gap there to take advantage of them during a retraining process. Moreover, we want to understand how to order an augmented dataset with adversarial inputs guided by metrics so that the retraining is efficient (with highest accuracy and using fewest inputs) for data scientists.} To the best of our knowledge, our study is the first one that applies metrics to guide a retraining using an augmented dataset with adversarial inputs in order to improve the model accuracy against adversarial inputs and keeping the trade-off with resource utilization, in order to obtain computational benefits, which is key as the retraining phase is time consuming.

In our study, we consider an augmented dataset with adversarial inputs obtained from the original training set using the FGSM method. Therefore our study provides the following novel contributions:

\begin{itemize}
    \item A comparison between state-of-the-art guidance metrics for a guided retraining.
        
    \item A comparison of three different configurations for doing a retraining against adversarial inputs. 
\end{itemize}

%\end{multicols}

\begin{figure*}[ht]
  \centering
  \includegraphics[width=.92\textwidth,height =0.6 \textheight,keepaspectratio]{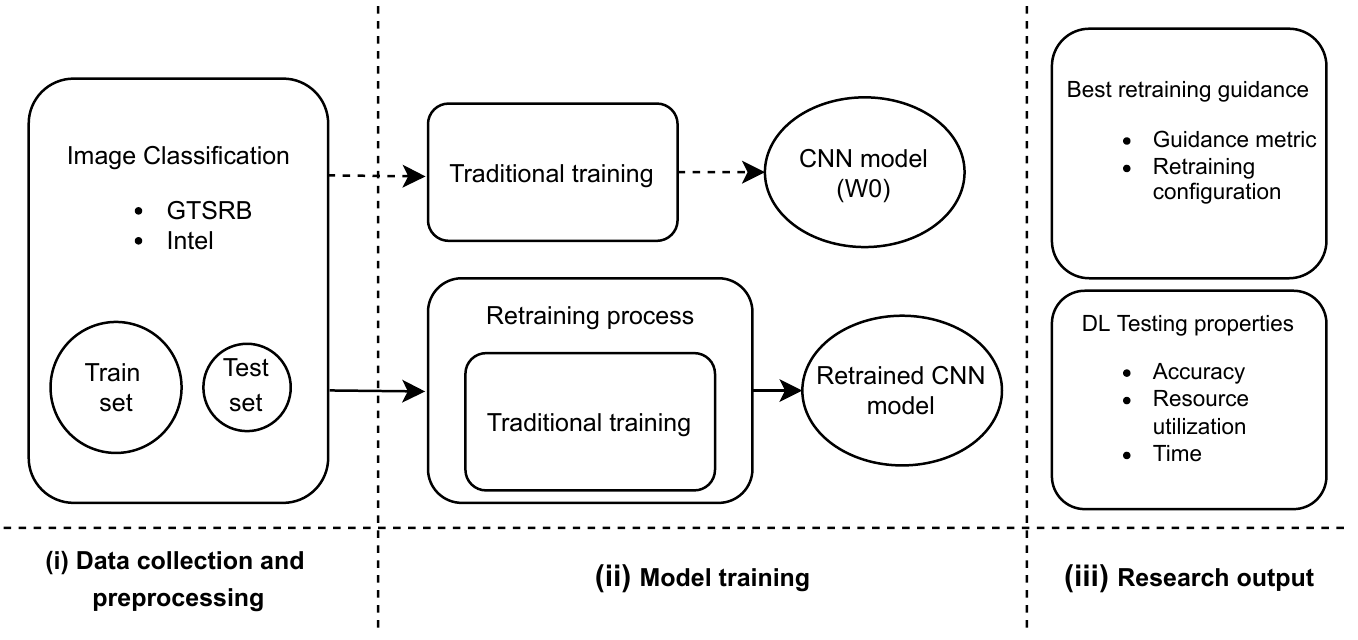}

  \caption{Schema of our empirical study %in the retraining process of a DL model, with three phases: (i) data collection and preprocessing phase, obtaining and preprocessing the data source; (ii) model training phase, training the DL model for solving the defined task (traditional training or realize the proposed retraining process); (iii) answering to the RQs.
}
  \label{diagram-1}
  \Description{Schema}
\end{figure*}

\begin{figure*}[t]
  \centering
  \includegraphics[width=.92\textwidth, height=0.6\textheight,keepaspectratio]{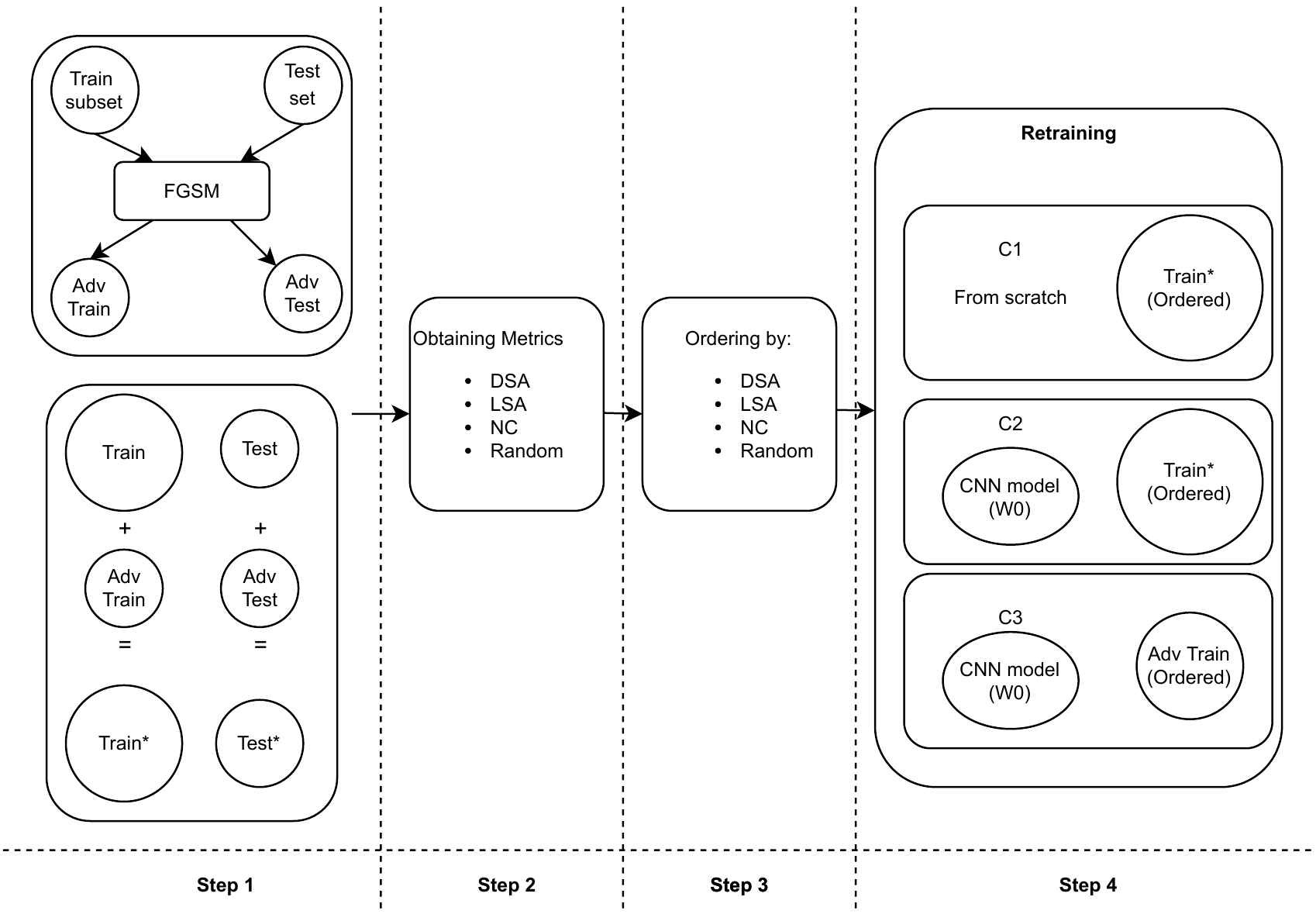}
  \caption{Schema of retraining process}
  \label{diagram_2}
  \Description{Schema}
\end{figure*}

%\begin{multicols}{2}

\section{Empirical study}\label{sec:empirical_study}

\subsection{Research questions}\label{sec:research questions}
The goal \cite{basili1994goal} of this research is to analyze \textit{guidance metrics and retraining configurations for a retraining process of CNN models} with the purpose of \textit{comparing them} with respect to \textit{DL testing properties such as accuracy, resource utilization and time} from the point of view of a \textit{data scientist} in the context of \textit{image classification}.

Thus, we aim at answering the following research questions (RQ):

\begin{itemize}
    \item RQ1 - Does the use of guidance metrics impact the accuracy, the resource utilization and the time required for the retraining of a CNN model?
    \item RQ2 - Does the configuration of the retraining of a CNN model impact the accuracy and the resource utilization required for retraining this model?
\end{itemize}

\subsection{Variables}

Table \ref{tab:variablestab} describes the variables of our study. With respect to the dependent variables, some details follow:

\begin{itemize}
  \item Accuracy, which measures the capability of the retraining phase of providing higher accuracy against adversarial inputs. We plan to measure the accuracy of the CNN models against an augmented test set, composed of the original test set and an adversarial test set obtained from the previous applying the FGSM to each one of the original test input.
  \item Resource utilization, which measures the input size to obtain the highest accuracy during the retraining phase.
  \item Time, which quantifies the time to compute each of the considered metrics for the corresponding dataset.
\end{itemize}

Another variable that influences our conclusions is the dataset, we experiment with two different datasets. Hence, we define the dataset as a nominal variable indicating the used dataset for training and retraining the models. They are from different domains and have different sizes.

\subsection{Study design}%\subsection{Experimental setup}

Figure \ref{diagram-1} shows the study design% and Figure \ref{diagram_2} the proposed retraining process
. First, the \emph{Data collection and preprocessing} which consists of the collection and preprocessing of raw data. Second, the \emph{Model training} which consists of the traditional training and proposed retraining phase. Finally, we provide answers to the RQs by analyzing our results w.r.t. DL testing properties. 

\subsubsection{Data collection and preprocessing: datasets}
\label{sec:experimental setup}

We evaluate the guidance metrics on CNNs using two multi-class, single-image classification datasets. First, the German Traffic Sign Recognition Benchmark (GTSRB), with 43 classes containing 39.208 unique images of real traffic signs in Germany.  It has been widely used in  DL research \cite{castanyer2021integration,stallkamp2012man,loukmane2020model}. This type of images are characterized by large changes of visual appearance for different causes (e.g., weather conditions). While humans can easily classify them, for DL systems (e.g., autonomous driving systems) still is a challenge. 

Second, the Intel Image Classification dataset with 6 classes containing 17.034 labeled images of natural scenes around the world, used in many studies as well \cite{wu2020xception,rahimzadeh2021wise,ren2022robustness}. It was provided by Intel corporation to create another benchmark in image classification tasks such as scene recognition \cite{Intel}. This scene recognition task is a daily task for humans and widely used application in industries such as tourism or robotics, still an emerging field for computer vision \cite{matei2020deep,sobti2021ensemv3x}.

The adversarial inputs are obtained using the FGSM method from the foolbox library \cite{rauber2017foolbox}. In the retraining the adversarial inputs are from the original training set (see ``Adv. Train'' in Fig. \ref{diagram_2} Step 1). Another adversarial set is obtained from the original test set (see ``Adv. Test'' in Fig. \ref{diagram_2} Step 1).

Using only one adversarial attack to create the adversarial inputs can be a limitation of our work. Nevertheless, when Goodfellow \emph{et al.} \cite{goodfellow2014explaining} presented FGSM, he stated that this type of adversarial examples can be generalized across architectures and training sets. Being aware of other attack methods, we chose this method because it is one of the most practical methods, and it is more widely used than other attack methods such as Basic Iterative Method (BIM), Jacobian-based Saliency Map Attack (JSMA), Projected Gradient Descent (PGD) or Carlini-Wagner (CW), compared to CW in particular, FGSM is more time efficient. On top of that, we focus our study in the retraining phase rather than creation of adversarial inputs. 

\subsubsection{Model training}

The traditional training of a model consists of using an available training set identified as ``Train'' in Figure \ref{diagram_2} and then to evaluate the model against a test set identified by ``Test'' in Figure \ref{diagram_2}. The traditional training is represented in the upper part of Fig. \ref{diagram-1} (ii) Model training phase. We plan to add in this phase a retraining process as it is represented in the lower part of Fig. \ref{diagram-1} (ii) Model training phase. This retraining uses adversarial inputs guided by the metrics and three different configurations to improve the original model, M,  against adversarial inputs. After obtaining a retrained model, M*, we evaluate the DL testing properties of this retrained model. The retraining process is shown more extensively in Figure \ref{diagram_2}. It comprises the following steps:

\begin{enumerate}
    \item \textbf{Step 1. Create adversarial inputs for training and testing:} We obtain two adversarial sets applying FGSM. First, to augment the training set, we apply FGSM to a subset of the original training set, ``Train'', to obtain the ``Adv. Train'' set, the number of the adversarial inputs created is from a small proportion of the entire ``Train'' set, thus we do not increase much the artificial inputs used in retraining and the naturalness is not diminished either. Putting these two sets together, we obtain ``Train*'' set. Second, to augment the original test set, ``Test'',  we apply FGSM to the entire set to obtain the ``Adv. Test'' set as it is too small compared to the augmented training set, ``Train*''. Putting these ``Test'' and ``Adv. Test'' sets together, we obtain ``Test*'' set.

    \item \textbf{Step 2. Obtain guidance metrics:} Based on the original trained model, M, the ``Train'' set and the new adversarial inputs for training ``Adv. Train'', we compute the different metrics for the augmented training set, ``Train*''.
    
    \item \textbf{Step 3. Order inputs w.r.t. the guidance metrics:} According to each metric, we order the inputs for the retraining. With this, we expect that the retrained model, M*, will be trained first with the images that are more difficult to classify and more informative according to the metrics' value.
    
    \item \textbf{Step 4. Retraining according to the configuration:} We implement the retraining in three different ways using ordered inputs according to the following configurations (see Figure \ref{diagram_2}):
    
    \begin{enumerate}
    \item \textbf{C1: Starting from scratch using the new adversarial inputs and original training set.} 
    %Configuration 1
    The new training set with the label ``Train*'' in Fig. \ref{diagram_2} from Step 1 is composed of the ``Train'' and ``Adv. Train'' sets. The model M* is retrained with this ``Train*'' set ordered by highest score of LSA, DSA, NC and Random, respectively, starting from scratch.
        
    %Configuration 2
    \item \textbf{C2: Starting from the original model M using the new adversarial inputs and original training set.} 
    The new training set is the same ``Train*'' set as in C1 with the difference that the model M* is retrained from the original model M weights.
    
    \item \textbf{C3: Starting from the original model M using only the new adversarial inputs.} 
    The new training set is only the ``Adv. Train'' set. The model M* is retrained with this set, also ordered by highest score of LSA, DSA, NC and Random, respectively, starting from the original model M weights.
        
    \end{enumerate}
\end{enumerate}

For each configuration, in the retraining phase, we execute a retraining of the models guided by the four metrics and for each metric we obtain 20 data points as shown in Figures \ref{fig:gtsrb_c1} - \ref{fig:intel_c3}. Each retraining run for each data point is computed from their respective initial weights. Also, we address randomness of resulted values by executing the retraining randomly, and not through incremental training.

\subsubsection{Data analysis}\label{sec:analysis}

\begin{figure*}[p]
  \centering
  \begin{subfigure}[b]{0.45\textwidth}
         \centering
         %\includegraphics[width=\textwidth]{gtsrb_c1_accuracy_both1} 
         %\includesvg[width=\textwidth]{gtsrb_c1_accuracy_both.ps}
         %\epsfbox[width=\textwidth]{gtsrb_c1_accuracy_both.eps}
         %\includegraphics[width=.92\textwidth,height =0.6 \textheight,keepaspectratio]
         \includegraphics[width=\textwidth]{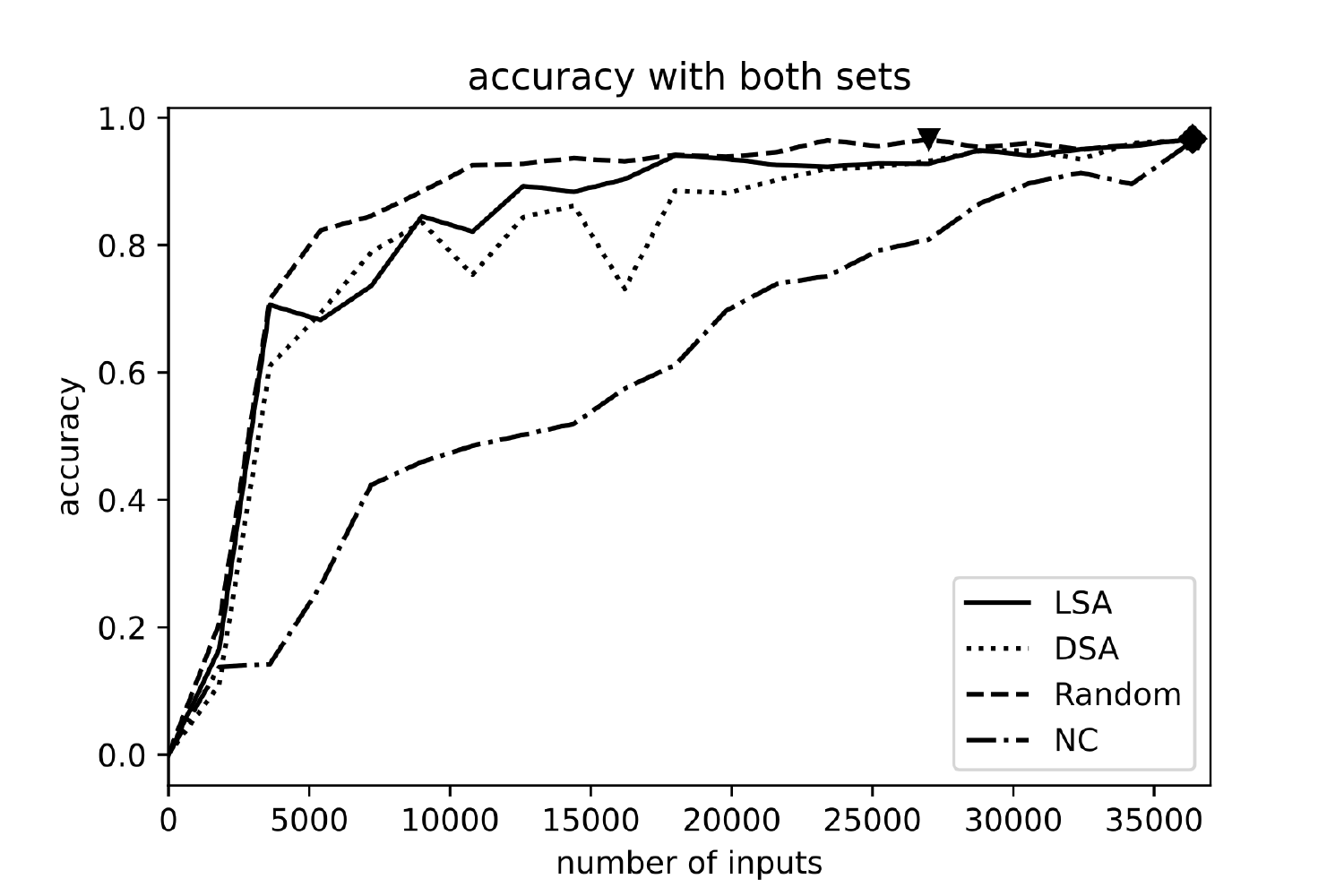}
         \caption{GTSRB dataset, C1.}
         \label{fig:gtsrb_c1}
    \end{subfigure}
  \hspace{1cm}
  \begin{subfigure}[b]{0.45\textwidth}
         \centering
         \includegraphics[width=\textwidth]{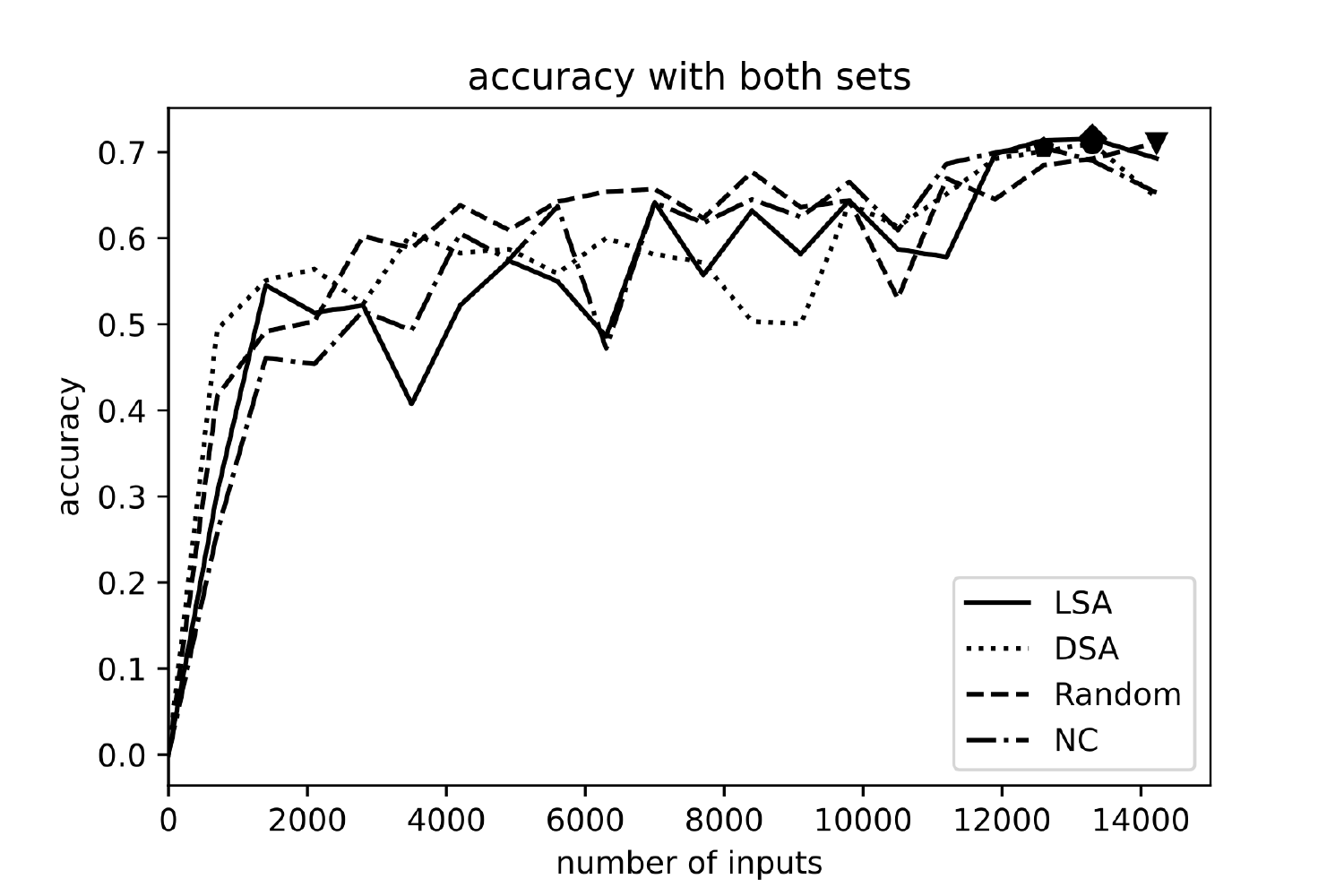}
         \caption{Intel dataset, C1.}
         \label{fig:intel_c1}
     \end{subfigure}

  \vspace{2.cm}
  \begin{subfigure}[b]{0.45\textwidth}
         \centering
         \includegraphics[width=\textwidth]{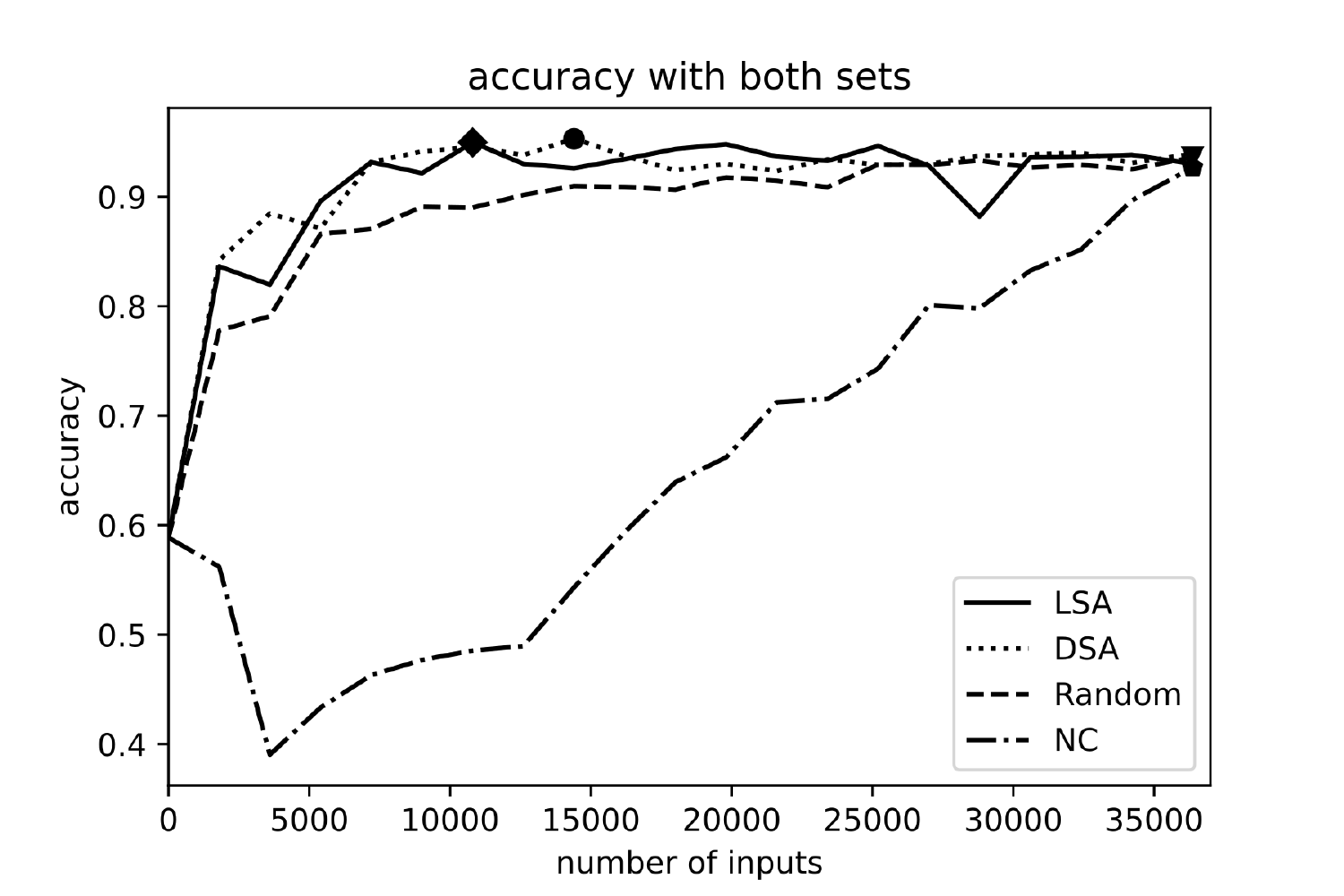}
         \caption{GTSRB dataset, C2.}
         \label{fig:gtsrb_c2}
     \end{subfigure}
  \hspace{1cm}
  \begin{subfigure}[b]{0.45\textwidth}
         \centering
         \includegraphics[width=\textwidth]{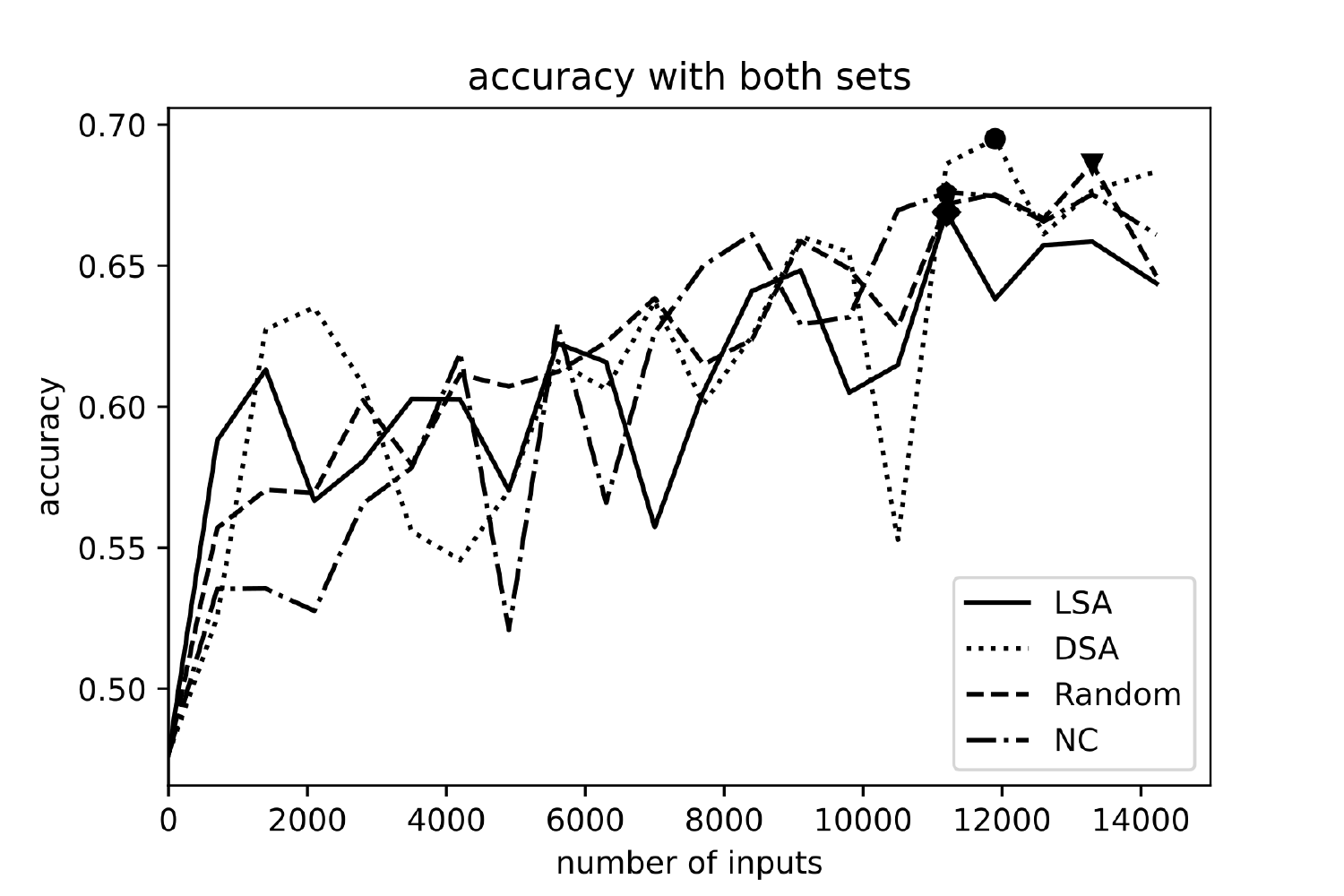}
         \caption{Intel dataset, C2.}
         \label{fig:intel_c2}
     \end{subfigure}

  \vspace{2.cm}

  \begin{subfigure}[b]{0.45\textwidth}
         \centering
         \includegraphics[width=\textwidth]{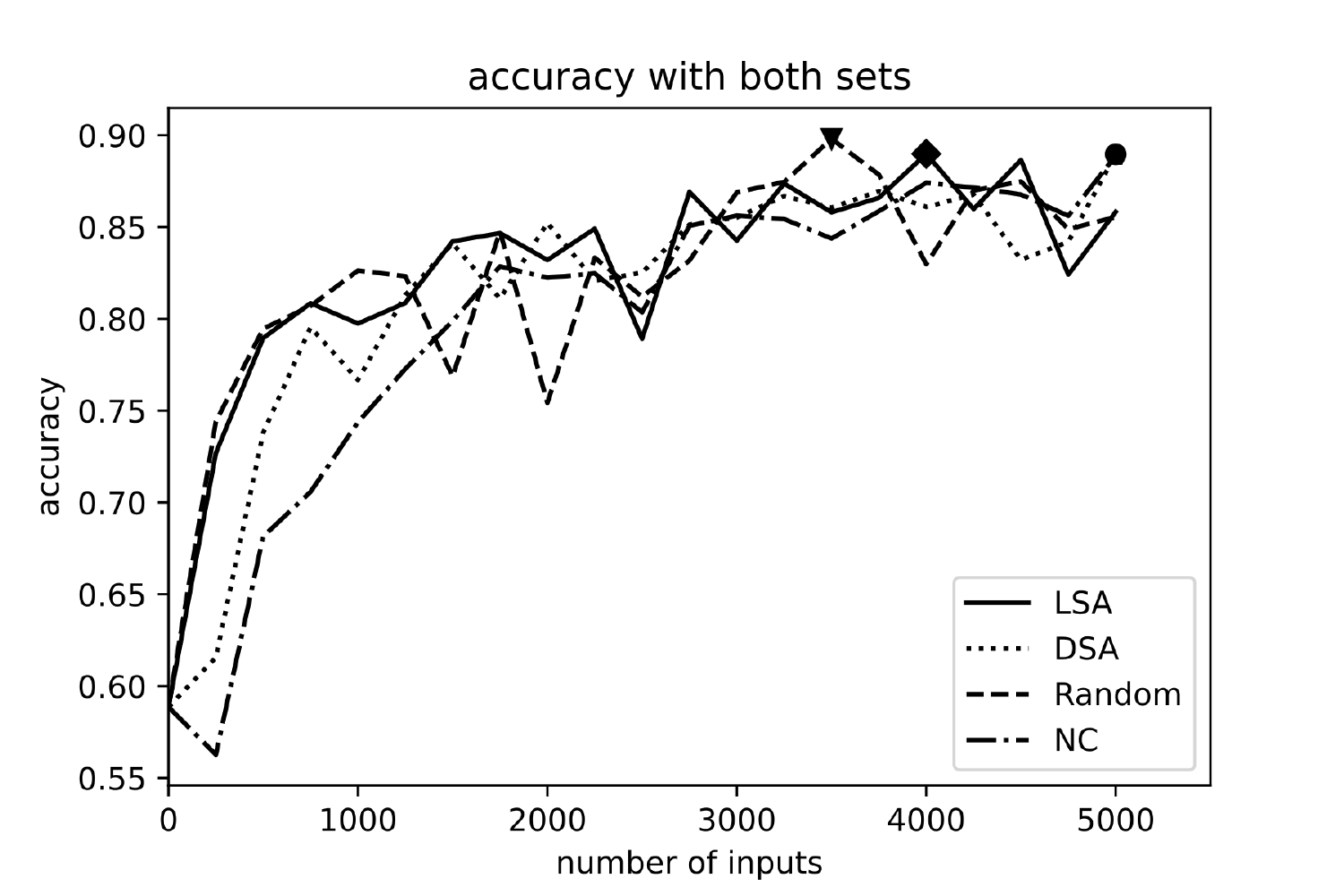}
         \caption{GTSRB dataset, C3.}
         \label{fig:gtsrb_c3}
     \end{subfigure}
  \hspace{1cm}
  \begin{subfigure}[b]{0.45\textwidth}
         \centering
         \includegraphics[width=\textwidth]{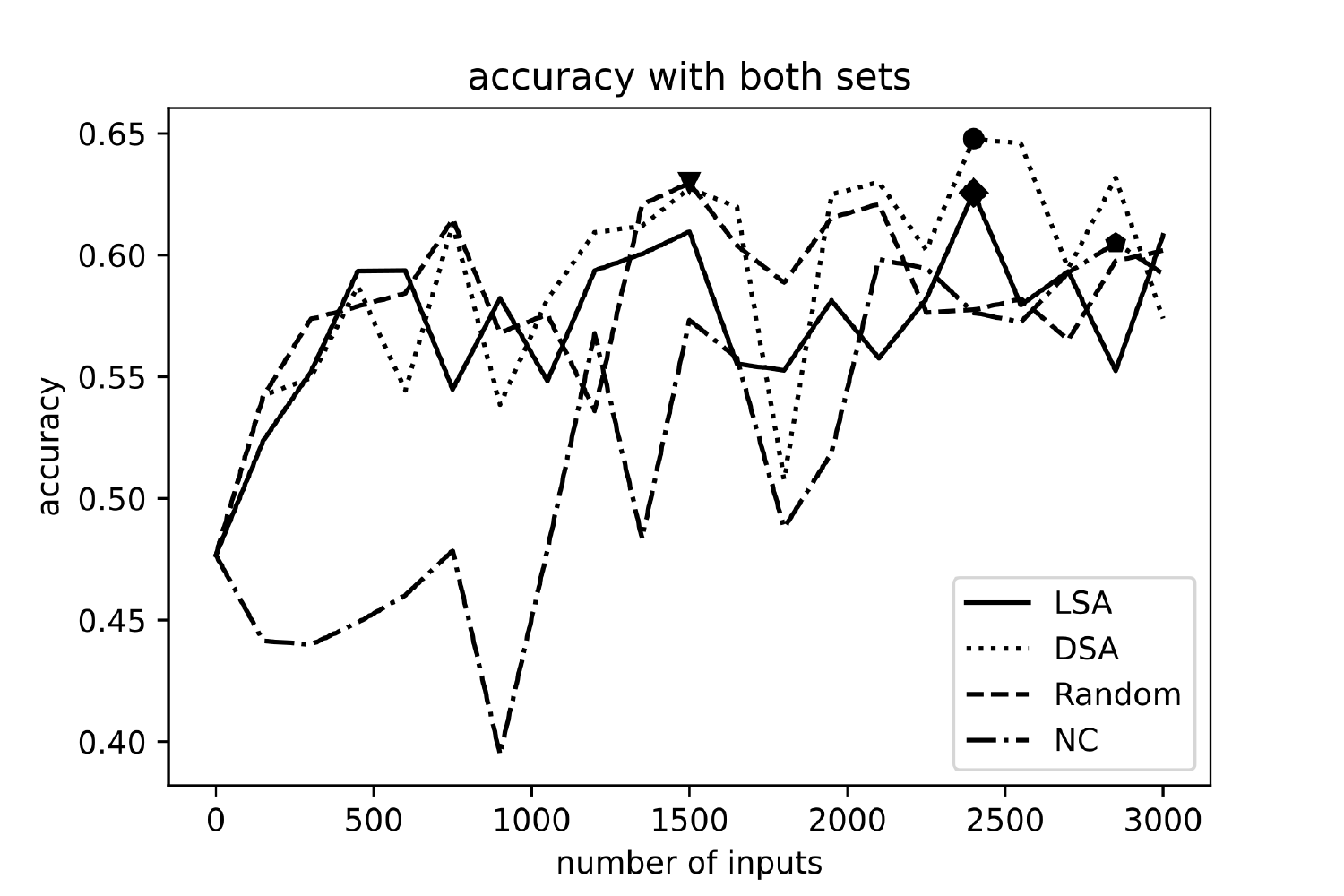}
         \caption{Intel dataset, C3.}
         \label{fig:intel_c3}
     \end{subfigure}

  \caption{Accuracy of the trained models.}
\end{figure*}

Figures \ref{fig:gtsrb_c1} - \ref{fig:intel_c3} correspond to the accuracy of the models against the augmented test set, ``Test*''. Table \ref{tab:RQ2_1} shows the accuracy against the same ``Test*'' set and resource utilization of the experiments  by dataset, configuration and metric used in each case. The ``Original accuracy'' column shows the accuracy of the original model M against this new augmented dataset, which is low due to the adversarial inputs, the ``Accuracy w.r.t. augmented test set'' column shows the highest accuracy during its respective retraining and the ``Resource utilization'' column shows the input size to obtain that highest accuracy. Table \ref{tab:time table} shows the time to compute the considered metrics for every input of the corresponding dataset.

In order to compare the impact that guidance metrics and retraining configurations have on the accuracy and resource utilization of the retrained models, we select the best model, according to the accuracy, during the retraining of the 20 data points for each combination of configuration and guidance metric (see marked points on Figures \ref{fig:gtsrb_c1} - \ref{fig:intel_c3} in Section \ref{sec:results}). We obtain the accuracy and resource utilization of these points and report these values in Tables \ref{tab:RQ2_1} - \ref{tab:RQ2_2} to determine if they can be improved by guiding the retraining with the studied metrics and using the studied retraining configurations. In addition to that, we evaluate the time it takes to compute the guidance metrics in Table \ref{tab:time table}.

We answer the RQs from two different angles. First, we compare the accuracy changes against the test set augmented with adversarial inputs in the 20 data points of the retrained models according to each configuration and guidance metric. The test set is composed of the original test set and adversarial inputs created with the same attack method but using inputs of the original test set; this way we ensure that the test data is never used during training or retraining. Second, we compare the input size required to obtain the model with highest accuracy between those data points according to each configuration and guidance metric. 

%%%%%%%%%%%%%%%%%%%%----

\begin{table*}[]
\captionsetup{justification=centering}
\caption{Accuracy against augmented test set (original test set and adversarial test set) and resource utilization when the model reach the maximum accuracy for C1, C2 and C3 applied to each dataset}
\label{tab:RQ2_1}
%\begin{adjustbox}{width=\columnwidth,center}
\scalebox{0.94}{%

\begin{threeparttable}
\begin{tabular}{|l|l|l|llll|llll|}

\hline
\textbf{}        & \textbf{}              & \textbf{}                  & \multicolumn{4}{l|}{\textbf{Accuracy w.r.t. augmented test set}}                                                                            & \multicolumn{4}{c|}{\textbf{Resource utilization}}                                                                                            \\ \hline
\textbf{Dataset} & \textbf{Config. \tnote{1}} & \textbf{Original accuracy} & \multicolumn{1}{l|}{\textbf{LSA}}   & \multicolumn{1}{l|}{\textbf{DSA}}   & \multicolumn{1}{l|}{\textbf{NC}} & \textbf{Random} & \multicolumn{1}{l|}{\textbf{LSA}}         & \multicolumn{1}{l|}{\textbf{DSA}}         & \multicolumn{1}{l|}{\textbf{NC}} & \textbf{Random}    \\ \hline
GTSRB            & C1                     & 0.589                      & \multicolumn{1}{l|}{\textbf{0.967}} & \multicolumn{1}{l|}{0.965}          & \multicolumn{1}{l|}{0.963}       & 0.966           & \multicolumn{1}{l|}{\textbf{36366/36366}} & \multicolumn{1}{l|}{36366/36366}          & \multicolumn{1}{l|}{36366/36366} & 27000/36366        \\ \hline
Intel            & C1                     & 0.477                      & \multicolumn{1}{l|}{\textbf{0.715}} & \multicolumn{1}{l|}{0.710}          & \multicolumn{1}{l|}{0.705}       & 0.710           & \multicolumn{1}{l|}{\textbf{13300/14224}} & \multicolumn{1}{l|}{13300/14224}          & \multicolumn{1}{l|}{12600/14224} & 14224/14224        \\ \hline
GTSRB            & C2                     & 0.589                      & \multicolumn{1}{l|}{0.950}          & \multicolumn{1}{l|}{\textbf{0.953}} & \multicolumn{1}{l|}{0.927}       & 0.936           & \multicolumn{1}{l|}{10800/36366}          & \multicolumn{1}{l|}{\textbf{14400/36366}} & \multicolumn{1}{l|}{36366/36366} & 36366/36366        \\ \hline
Intel            & C2                     & 0.477                      & \multicolumn{1}{l|}{0.669}          & \multicolumn{1}{l|}{\textbf{0.695}} & \multicolumn{1}{l|}{0.676}       & 0.686           & \multicolumn{1}{l|}{11200/14224}          & \multicolumn{1}{l|}{\textbf{11900/14224}} & \multicolumn{1}{l|}{11200/14224} & 13300/14224        \\ \hline
GTSRB            & C3                     & 0.589                      & \multicolumn{1}{l|}{0.890}          & \multicolumn{1}{l|}{0.890}          & \multicolumn{1}{l|}{0.889}       & \textbf{0.898}  & \multicolumn{1}{l|}{4000/5000}            & \multicolumn{1}{l|}{5000/5000}            & \multicolumn{1}{l|}{5000/5000}   & \textbf{3500/5000} \\ \hline
Intel            & C3                     & 0.477                      & \multicolumn{1}{l|}{0.626}          & \multicolumn{1}{l|}{\textbf{0.648}} & \multicolumn{1}{l|}{0.605}       & 0.630           & \multicolumn{1}{l|}{2400/3000}            & \multicolumn{1}{l|}{\textbf{2400/3000}}   & \multicolumn{1}{l|}{2850/3000}   & 1500/3000          \\ \hline

\end{tabular}
\begin{tablenotes}
    \item[1] Configuration of retraining.
    \item[] Note: bold numbers indicate the accuracy and resource utilization for the model with highest accuracy during retraining. 
\end{tablenotes}
\end{threeparttable}
}

%\end{adjustbox}
\end{table*}

\begin{table*}[]
\captionsetup{justification=centering}
\caption{Accuracy against augmented test set (original test set and adversarial test set) and resource utilization when the model reach the maximum accuracy for C2 (data point with same number of inputs used in C3) and C3 applied to each dataset}
\label{tab:RQ2_2}
%\begin{adjustbox}{width=\columnwidth,center}
\scalebox{0.94}{
\begin{tabular}{|l|l|l|llll|llll|}
\hline
\textbf{}        & \textbf{}              & \textbf{}                  & \multicolumn{4}{l|}{\textbf{Accuracy w.r.t. augmented test set}}                                                                            & \multicolumn{4}{c|}{\textbf{Resource utilization}}                                                                                          \\ \hline
\textbf{Dataset} & \textbf{Config.} & \textbf{Original accuracy} & \multicolumn{1}{l|}{\textbf{LSA}}   & \multicolumn{1}{l|}{\textbf{DSA}}   & \multicolumn{1}{l|}{\textbf{NC}} & \textbf{Random} & \multicolumn{1}{l|}{\textbf{LSA}}        & \multicolumn{1}{l|}{\textbf{DSA}}        & \multicolumn{1}{l|}{\textbf{NC}} & \textbf{Random}    \\ \hline
GTSRB            & C2                     & 0.589                      & \multicolumn{1}{l|}{\textbf{0.860}} & \multicolumn{1}{l|}{\textbf{0.860}} & \multicolumn{1}{l|}{0.400}       & 0.820           & \multicolumn{1}{l|}{\textbf{5000/36366}} & \multicolumn{1}{l|}{\textbf{5000/36366}} & \multicolumn{1}{l|}{5000/36366}  & 5000/36366         \\ \hline
Intel            & C2                     & 0.477                      & \multicolumn{1}{l|}{0.590}          & \multicolumn{1}{l|}{\textbf{0.595}} & \multicolumn{1}{l|}{0.560}       & 0.590           & \multicolumn{1}{l|}{3000/14224}          & \multicolumn{1}{l|}{\textbf{3000/14224}} & \multicolumn{1}{l|}{3000/14224}  & 3000/14224         \\ \hline
GTSRB            & C3                     & 0.589                      & \multicolumn{1}{l|}{0.890}          & \multicolumn{1}{l|}{0.890}          & \multicolumn{1}{l|}{0.889}       & \textbf{0.898}  & \multicolumn{1}{l|}{4000/5000}           & \multicolumn{1}{l|}{5000/5000}           & \multicolumn{1}{l|}{5000/5000}   & \textbf{3500/5000} \\ \hline
Intel            & C3                     & 0.477                      & \multicolumn{1}{l|}{0.626}          & \multicolumn{1}{l|}{\textbf{0.648}} & \multicolumn{1}{l|}{0.605}       & 0.630           & \multicolumn{1}{l|}{2400/3000}           & \multicolumn{1}{l|}{\textbf{2400/3000}}  & \multicolumn{1}{l|}{2850/3000}   & 1500/3000          \\ \hline
\end{tabular}
}
%\end{adjustbox}
\end{table*}

\begin{table}[]
\captionsetup{justification=centering}
\caption{Time in hours (hh:mm:ss) to obtain the respective values}
\label{tab:time table}

\begin{tabular}{|l|l|l|l|l|}
\hline
\textbf{Dataset} & \textbf{LSA values} & \textbf{DSA values} & \textbf{NC values} & \textbf{Random} \\ \hline
GTSRB   & 00:01:35      & 00:24:54      & 03:17:38         & 00:00:00  \\ \hline
Intel   & 00:00:57      & 00:02:04      & 01:08:18         & 00:00:00  \\ \hline
\end{tabular}

\end{table}

\section{Results}\label{sec:results}

In this section, we report the results of our empirical study, answering RQ1 and RQ2, and highlighting key takeaways.

To better describe the results, we use the Figures \ref{fig:gtsrb_c1} - \ref{fig:intel_c3} and Tables \ref{tab:RQ2_1}, \ref{tab:RQ2_2} and \ref{tab:time table}, as explained in Section \ref{sec:analysis}. 

\subsection{Does the use of guidance metrics impact the accuracy, the resource utilization and the time required for the retraining of a CNN model? (RQ1)}

\textit{Guidance metrics and accuracy.}
According to the Figures \ref{fig:gtsrb_c1} - \ref{fig:intel_c3}, we found in the experiments that SA metrics  have the best selection of inputs for the retraining as stated by the observed accuracy. Five out of six guided retraining runs obtained the best accuracy using SA metrics (highlighted with bold font in Table \ref{tab:RQ2_1}).

\textit{Guidance metrics and resource utilization.}
According to the results of using C2 with the GTSRB dataset, the impact of the method can be noticed with greater difference. Clearly, SA metrics can reach a better model with less inputs, as much as 14.400/36.366 inputs in the best case with 0.953 of accuracy, as shown in Figure \ref{fig:gtsrb_c2} and Table \ref{tab:RQ2_1}. On the other hand, this is not so clear when using Intel dataset, which may be due to the size of the dataset: as there are not enough inputs, the metrics cannot make a difference with just a percentage of an already relatively small dataset. Nevertheless, DSA was the best option according to Figure \ref{fig:intel_c2}.

Overall, SA metrics may be the best due to the fact that they first identify the inputs that are most different from the previous training inputs, so the model learns features that are different from those previously learned, in order to correctly classify adversarial inputs. On the other hand, for NC metric it may be more difficult to identify good inputs for retraining because this metric identifies inputs that cover more neurons according to certain thresholds, which in retraining may not be significant.

\textit{Guidance metrics and time.} We obtained the time to calculate the metrics as stated in Table \ref{tab:time table}. To obtain NC values it takes more than 7x in terms of time w.r.t. the time needed to obtain SA values, which may be due to the library used and to a non-optimized metric computation. In addition, the lower time required to get SA values may be also due to the size of the datasets, as it is known that the computation of SA metrics tends to quickly augment \cite{weiss2021review}. We can observe this trend when comparing the ratio of NC time to DSA time in each dataset. When we consider Intel dataset, computing NC values takes 34x the DSA time, but when using GTSRB dataset, this ratio reduces to almost 8x.

\begin{tcolorbox}
    Key takeaways for RQ1:
    \begin{itemize}
        \item SA metrics  have the best selection of inputs for the retraining as stated by the observed accuracy.
        \item SA metrics can reach a better model with less inputs.
        \item To obtain NC values it takes more than 7x in terms of time w.r.t. the time needed to obtain SA metrics.
    \end{itemize}
\end{tcolorbox}

\subsection{Does the configuration of the retraining of a CNN model impact the accuracy and the resource utilization required for retraining this model? (RQ2)}

\textit{Configuration of the retraining and accuracy.} According to the total number of used inputs, as expected, C1 and C2 reached higher values of accuracy than C3. Nevertheless, between these two configurations, the benefits of using C2 are greater because with less inputs, nearly less than a half in the best case (see Table \ref{tab:RQ2_1}), it is possible to complete the retraining with high accuracy. Due to the initial weights used in C2, this incremental training creates bias towards the original training set, which can be the reason why C2 does not worsen its accuracy against inputs from the original test set: the help of the adversarial inputs ``Adv. Train'' that augmented the dataset makes this model better against the adversarial inputs from the augmented test set ``Test*''.

\textit{Configuration of the retraining and resource utilization.}
According to the resource utilization, C2 is the best option of the studied configurations, nevertheless, using C3 shows some benefits. Table \ref{tab:RQ2_2} shows the same values of Table \ref{tab:RQ2_1} for C3, but for C2 the accuracy and resource utilization are obtained in an specific data point, when the model has only used the same input size for C3 (e.g., 5000 inputs when using GTSRB and 3000 inputs when using Intel dataset). Although C3 was expected to have lower accuracy than the other two configurations, because of the available input size for retraining, Table \ref{tab:RQ2_2} shows that C3 can be a good option if we want to execute a retraining with just few inputs. The values in the rows with accuracy of C3 for both datasets are greater in their respective values for C2.

We can compare C3 with C2 because the starting point is the same original model M and we are executing a retraining with the same number of inputs, but for C2 the inputs are chosen from the entire training set and adversarial set (``Train*''), while for C3 the inputs are only from the adversarial set (``Adv Train'').

\begin{tcolorbox}
    Key takeaways for RQ2:
    \begin{itemize}
        \item The benefits of using C2 are greater because with less inputs, nearly less than a half in the best case, it is possible to execute a retraining obtaining high accuracy.
        \item  C3 can be a good option for the retraining in cases in which we prioritize resource utilization.
    \end{itemize}
\end{tcolorbox}

%\end{multicols}

%\begin{multicols}{2}
\section{Discussion}\label{sec:discussions}

Our experimental results focused on the retraining of models against adversarial inputs, taking into consideration the most used and practical state-of-the-art metrics: LSA, DSA, NC and Random. On top of that, we considered the manner the retraining is done.

As observed in the results, we have verified that NC is not a consistent metric to take into consideration to select inputs when doing retraining. Previous studies, mainly Harel-Canada et al. \cite{harel2020neuron}, have already stated that NC should not be trusted as a guidance metric for DL testing. On the one hand, NC measures the proportion of neurons activated in a model and it assumes that increasing this proportion improves the quality of a test suit. But on the other hand, Harel-Canada et al. showed that increasing NC value leads to fewer defects detected, less natural inputs and more biased prediction preferences, which is contrary to what we would expect when selecting inputs for a retraining process to have a better model and reduce the input size for that. In our study, the results confirm that NC should not be trusted as a guidance metric. Furthermore, Table \ref{tab:time table} shows an evident  disadvantage on time when computing NC. Overall, different metrics rather than NC should be used when doing guided retraining.

Ma \emph{et al.} \cite{ma2021test} found that uncertainty-based and surprise-based metrics are the best at selecting retraining inputs and lead to improvements in the number of used inputs, up to twice faster than random selection, in order to find an answer on how to select additional training inputs to improve classification accuracy. Regarding SA metrics, we confirm the results obtained in that study: SA metrics compared to the baseline of random selection and also to NC, are better and lead to faster improvements with a satisfactory model. We consider this work as a complement to the results of the RQ about the selection for retraining of Ma \emph{et al.}, as we focused our work on increment the model’s accuracy against adversarial inputs. 
Although our results show that these metrics improve the model faster (in terms of the input size used for retraining), we have found that the size of the dataset greatly affects the results, because when using small datasets it is not as fast. Ma \emph{et al.} only considered datasets with more than 50.000 available inputs to use in the retraining (the entire training set), but in real world applications is not always possible to have such large datasets. Unlike this work, we have experimented with a smaller dataset in which the improvements are minor (Intel dataset). Previous studies, our results of SA metrics and new implementations of these metrics \cite{weiss2021review,ouyang2021corner,kim2020reducing} aim to be a baseline for test methods in DL testing: for data selection in retraining processes, data generation, data selection, etc.

Additionally, an important finding is the variable that we also considered: configuration of retraining. All the related works have only experimented with their own way to execute the retraining and even sometimes they are not explicit with how they did the retraining process. As we observe in the results, the configuration can change the performance of the model considerably. We have uncovered a new challenge on finding efficient configurations to retrain deep learning models. Considering the studied configurations, using the best configuration studied in this work (C2) can give data scientists a fast and efficient method of retraining.

When using the combination of DSA metric and C2 retraining configuration, we observe that it can lead to results that are aligned with green AI research, which refers to research that takes into account computational costs to reduce the computational resources spent \cite{schwartz2020green, castanyer2021design}, as  less inputs are needed for retraining and also less time to compute the values of the metric w.r.t. the evaluated options for both independent variables. We observe greater benefits of using C2 over C3, as we give greater weight to the efficiency without diminishing model's accuracy, and encourage data scientists to build greener DL models.
 
\section{Threats to validity}\label{sec:threats}

In this section, we report the limitations of our empirical study and some mitigation actions taken to reduce them as much as  possible.

Regarding construct validity, we include two original models using CNN architecture trained in two different datasets, respectively, in order to mitigate mono-operation bias. Derived from the metrics selected, we use LSA, DSA, NC and Random metrics to guide the retraining runs, which may have potential threats. However, these metrics have been used in previous work as shown in Section \ref{sec:related work}, lowering this threat. And according to threats related to the configurations, we have reviewed the relevant literature and searched for retraining configurations using adversarial inputs.

Concerning conclusion validity, the quality of the DL models and implementations depend on the experience of the developers. To mitigate this, we provide the implementations organized and following the  ``Cookiecutter Data Science'' project structure\footnote{https://drivendata.github.io/cookiecutter-data-science}, making them as simple as possible. To increase reliability in our study, we detail the procedure to reproduce our work: the process is shown in Section \ref{sec:empirical_study}, datasets and replication package are available online. Also, we address the randomness of our results by starting the retraining runs for each data point from their respective initial weights (from original model weights for C2 and C3, and from scratch for C1).

Two threats to internal validity are the implementation of the
studied DL models, as well as the computation of the metrics. We
used available replication packages from the authors of the metrics,
using the same configurations they used for the experiments to
minimize this risk. Also, different models are used with different datasets, mitigating that the results of our study of guidance metrics and configurations are caused accidentally.

Threats to external validity stem mainly from the number of datasets, models and the adversarial generation algorithm considered. Our results depend also on the datasets, type of architecture considered and the device used for training. We believe our results are applicable to image classification datasets. Some of these threats are addressed by the use of two image datasets, several state-of-the-art metrics and an adversarial attack widely used by scientific community. Regarding the architecture type, as the adversarial inputs can be generalized across different of these \cite{goodfellow2014explaining} and we use them for the retraining, we also believe that results for other architectures could be similar to the results for CNN architecture, only further experiments can reduce this threat.

%\end{multicols}

\section{Conclusions and Future Work}\label{sec:conclusions}
In this work we have studied DL testing metrics for guiding retraining of models. We performed an empirical study with the metrics and also considered three different configurations of retraining against adversarial inputs and did a comparison of the metrics and the configurations. In summary, we observe that \emph{(i)} the models are increasing their accuracy against the test set augmented with adversarial inputs as it was sought in the objectives of this work, and \emph{(ii)} there are computational benefits of using certain metrics and configurations.

The empirical study showed that the SA metrics (such as LSA and DSA) as guidance for a retraining phase are useful for data scientists when using the following configuration: an augmented training dataset with adversarial inputs, starting from original model weights.

With the previous configuration and metrics, we can improve the accuracy of the models against adversarial inputs by up to 61.8\% on the GTSRB dataset and up to 45.7\% on the Intel dataset without the need of using many inputs. Therefore, this can be done by using 39.6\% of the inputs on the former dataset and 83.7\% of  the inputs on the latter when using DSA. Using random can guide to similar levels of accuracy but when using the recommended configuration, the computational benefits of using SA are that less inputs are required as discussed above in percentage. However, we do not recommend the use of NC metric, and prevent that  when considering another configuration such as: using an augmented dataset with adversarial inputs and starting from scratch, can be time consuming, as almost 100\% of the inputs need to be used to obtain similar accuracy to the recommended configuration. 

Additionally, we revealed that the size of the dataset is important when implementing the recommended metrics and configuration. Taking this into account, we need to assess whether it would be worth calculating the metrics when using only small datasets.

In the next step, the use of other adversarial attacks for the creation of adversarial inputs and reproduction of our experiments on different datasets of varied sizes, unbalanced datasets and also non-images datasets \cite{kim2021multimodal} are required to generalize our findings. Particularly, experiments with other DL architectures are required to confirm or reject that our findings can be used across different architectures. Also, our results should be compared to guided retraining runs using new refinements of SA \cite{kim2020reducing,weiss2021review,ouyang2021corner} and other testing metrics such as uncertainty-based metrics.  

\section*{Data availability}
Following open science principles, the replication package with data and implementation can be found on %\url{https://github.com/fjdurlop/guided-retraining} and 
\url{https://doi.org/10.5281/zenodo.5904550}.

\begin{acks}
This work was partially supported by: the ``UNAM-DGECI: Iniciación a la Investigación (verano-otoño 2021)'' scholarship provided by Universidad Nacional Autónoma de México (UNAM); the DOGO4ML Spanish research project (ref. PID2020-117191RB-I00); the ``Beatriz  Galindo''  Spanish Program BEAGAL18/00064; the Austrian Science Fund (FWF): I 4701-N; and, the project Continuous Testing in Production (ConTest) funded by the Austrian Research Promotion Agency (FFG): 888127.
\end{acks}

\clearpage

\bibliographystyle{ACM-Reference-Format}
\bibliography{references}

%
%Table 1 presents the analyzed studies in chronological order. Many studies improve the retraining of models or analyze different testing methods for test adequacy criterion.
%

\end{document}